\documentclass[preprint2]{aastex}
\usepackage{psfig}

\newcommand{\NH}{\mbox{$N_{\rm H}$}}             
\newcommand{\hi}{H{\small ~I}}                   
\newcommand{\hii}{H{\small ~II}}                 
\newcommand{\lsim}{\hbox{$_{\normalsize \sim}^{\normalsize <}$\ }} 

\slugcomment{Accepted by The Astronomical Journal}

\shorttitle{M101 Hypernova Remnant Candidates}
\shortauthors{Snowden et al.}

\begin{document}

\title{Reconsidering the Identification of M101 Hypernova Remnant Candidates}

\author{S. L. Snowden\altaffilmark{1,}\altaffilmark{2}, 
K. Mukai\altaffilmark{1,}\altaffilmark{3}, and W. Pence\altaffilmark{4}}
\affil{Code 662, NASA/Goddard Space Flight Center, Greenbelt, MD 20771}

\and

\author{K. D. Kuntz\altaffilmark{5}}
\affil{Joint Center for Astrophysics, University of Maryland Baltimore County,
Baltimore, MD, 21250}

\altaffiltext{1}{Universities Space Research Association}
\altaffiltext{2}{E-mail-I:snowden@riva.gsfc.nasa.gov}
\altaffiltext{3}{E-mail-I:mukai@milkyway.gsfc.nasa.gov}
\altaffiltext{4}{E-mail-I:pence@milkyway.gsfc.nasa.gov}
\altaffiltext{5}{E-mail-I:kuntz@halo.gsfc.nasa.gov}

\begin{abstract}

Using a deep {\it Chandra} AO-1 observation of the face-on spiral galaxy 
M101, we examine three of five previously optically-identified 
X-ray sources which are spatially correlated with optical supernova 
remnants (MF54, MF57, and MF83).  The X-ray fluxes from these objects, 
if due to diffuse emission from the remnants, are bright enough to 
require a new class of objects, with the possible attribution 
by Wang to diffuse emission from hypernova remnants.  
Of the three, MF83 was considered the most likely candidate for such an 
object due to its size, nature, and close positional coincidence. 
However, we find that MF83 is clearly ruled out as a hypernova remnant by 
both its temporal variability and spectrum.  The bright X-ray sources 
previously associated with MF54 and MF57 are seen 
by {\it Chandra} to be clearly offset 
from the optical positions of the supernova remnants by several arc seconds, 
confirming a result suggested by the previous work.  MF54 does have a faint 
X-ray counterpart, however, with a luminosity and temperature consistent 
with a normal supernova remnant of its size.
The most likely classifications of the sources are as X-ray binaries.
Although counting statistics are limited, over the 0.3--5.0 keV
spectral band the data are well fit by simple absorbed power laws with
luminosities in the $10^{38}-10^{39}$~ergs~s$^{-1}$ range.
The power law indices are softer than those of Milky Way LMXB of
similar luminosities, and are more consistent with those of the
Large Magellanic Cloud.  Both the high luminosity and the soft
spectral shape favor these being accreting black hole binaries
in high soft states.

\end{abstract}


\section{Introduction}

M101 (NGC 5457), a face-on Sc galaxy at a distance of 7.2~Mpc \citep{sea98}, 
is an ideal subject
for the study of spiral galaxies of types similar to the Milky Way.  The 
low Galactic column density ($\sim1.2\times10^{20}$ HI cm$^{-2}$) allows 
investigations of X-ray emission down to energies of $\sim{1\over4}$~keV. 
It is relatively large with a $D_{25}$ diameter of $23\farcm8$,
$\sim50$~kpc, compared to $\sim24$~kpc for the Milky Way.  With the 
arc-second resolution of the {\it Chandra X-ray Observatory} 
\citep{wos96}, structures as small as $\sim30$~pc
can be resolved, and their positions can be well determined.

By correlating a catalog of optically-identified supernova remnants 
\citep{mf97} in M101 with X-ray sources detected in a {\it ROSAT} HRI 
deep observation \citep{wea99}, \citet{w99} 
identified five positional coincidences.  Because of the 
implied high luminosity of the objects, $\sim10^{38}-10^{39}$ ergs s$^{-1}$, 
if the observed emission was due to shock-heated interstellar gas then a 
new class of objects was required as the input energy would
be well beyond that of a normal supernova.  Wang speculated that the
excess energy could be due to the progenitor events being hypernovae, 
the name proposed by \citet{p98} for the entire gamma-ray burst (GRB) 
and afterglow event.  In a hypernova, the GRB ejecta (e.g., the outer 
material of a star that can be ejected during the period when a compact 
object enters its atmosphere preceding the object/helium-core collapse) 
collides with the ambient medium surrounding the event (\citet{p98} and 
references within).  This can produce an afterglow which can be hundreds 
of times more luminous than that of a normal supernova.

\begin{deluxetable}{cccccccc}
\tabletypesize{\scriptsize}
\tablecaption{Source Positions. 
\label{tbl:hyper}}
\tablewidth{0pt}
\tablehead{
\colhead{Supernova} & \multicolumn{3}{c}{Matonick \& Fesen (1997)}
& Wang (1999) & \multicolumn{3}{c}{Pence et al. (2001)} \\
\cline{2-4} \cline{6-8} \\
\colhead{Remnant}   & \colhead{R.A. (J2000)} & \colhead{Decl. (J2000)}
& \colhead{Diameter\tablenotemark{a}} & \colhead{Source} & \colhead{Source}
& \colhead{R.A. (J2000)} & \colhead{Decl. (J2000)}}
\startdata
MF54\tablenotemark{b} & 14 03 20.7 & 54 19 42.2 & 20
   & H29 & P70 & 14 03 21.5 & 54 19 45.9 \\
MF54\tablenotemark{c} & 14 03 20.7 & 54 19 42.2 & 20
   & $-$ & P67 & 14 03 20.6 & 54 19 42.3 \\
MF57 & 14 03 24.2 & 54 19 44.0 &  30 & H30 & P76  & 14 03 24.2 & 54 19 48.9 \\
MF83 & 14 03 35.9 & 54 19 24.1 & 200 & H36 & P104 & 14 03 36.0 & 54 19 24.8 \\
\enddata
\tablenotetext{a}{Diameter of the supernova remnant in parsecs.}
\tablenotetext{b}{Bright source likely detected by Wang et al. (1999).}
\tablenotetext{c}{{\it Chandra} source likely associated with MF54.}
\end{deluxetable}

In this paper we examine the three X-ray source/optical supernova 
remnant correlations which were considered as possible hypernova remnant 
candidates by \citet{w99}, and which were included in the field covered by 
the S3 (backside illuminated) CCD chip during our deep ($\sim100$~ks) 
{\it Chandra} Advanced CCD Imaging Spectrometer (ACIS, \citet{gea92};  
\citet{bea98}) observation of M101 (Table~\ref{tbl:hyper}).  
All of these objects have
luminosities substantially higher than that expected from a normal
supernova remnant implying that if the fluxes originate from diffuse 
emission the progenitor events were extremely energetic.  Of these three 
objects, the X-ray counterpart of MF83 (the MF nomenclature refers to the
source number in \citet{mf97}) was deemed the most likely to be 
a hypernova remnant and was discussed most extensively by \citet{w99}. 
(Unfortunately, the other more likely hypernova remnant candidate in
M101, NGC~5471B, lies outside of the region covered by the S3 chip, 
or by any operational CCD during the observation.)  The other two 
objects, MF54 and MF57, were deemed less plausible as hypernova remnants 
as their optical morphology was filled (rather than shell-like) and thus 
not likely to have entered the X-ray bright pressure-driven snowplow
phase.  In addition, the angular offsets between the positions of the 
optical and X-ray counterparts were also larger for MF54 and MF57 than
for MF83, and were border-line cases for association.  We will discuss 
all three sources in this paper.  Section~\ref{sec:data} discusses the 
data and data analysis, \S~\ref{sec:discussion} discusses the results,  
and \S~\ref{sec:results} presents our conclusions.

\section{Data and Analysis}
\label{sec:data}

The new data presented here are from a deep (97.4~ks) {\it Chandra} AO-1 
observation of M101 (Sequence ID 600111) which took place on 2000 March 26-27.  
The observation primarily utilized the ACIS S3
backside-illuminated CCD chip to maximize the instrumental sensitivity 
at low energies.  The pointing direction 
($\alpha,\delta_{2000}=14^{\rm hr}03^{\rm m}7.2^{\rm s},54^\circ21'35''$)
was slightly offset from the galactic nucleus 
($\alpha,\delta_{2000}=14^{\rm hr}03^{\rm m}12^{\rm s},54^\circ20'59''$)
to provide better S3 coverage of the region of M101 exhibiting diffuse
$1\over4$~keV emission \citep{sp95}, as well as other objects of interest.  
The observation occurred during a quiet period of the particle background 
providing an exceptionally clean data set.  A more extensive discussion of 
the observation will be provided in \citet{pea01}, which will provide a
catalog of the detected sources.

Processing the delivered data consisted of executing the ``acisclean''
perl script of K. Arnaud and I. George (private communication) to remove 
bad pixels and provide event grade selection.  Appropriate detector 
response matrices (rmfs) and effective 
area files (arfs) were produced for the three sources, again using scripts 
provided by K. Arnaud along with the CIAO software package of the 
{\it Chandra} X-ray Center.  Source and background data were extracted 
and spectral and temporal properties fit using the HEASARC HEASoft package.
The sources were all reasonably bright with source counts ranging from
$\sim350-1000$.  

\subsection{Temporal Variability}

As extended objects, hypernovae remnants should show no temporal variation
over the duration of the observation.  To search for any variability, we
binned the data into 2~ks intervals to provide for reasonable statistics 
and fit for a constant count rate.  The data for P70 (MF54/H29) and P76 
(MF57/H30) were consistent with a constant flux with reduced $\chi^2$ 
values of 0.92 and 0.76, respectively.  (The source names P67, P70, P76,
and P104 used in this paper refer to the source numbers in \citet{pea01},
while the source names H29 and H30 refer to the source numbers in 
\citet{wea99}.

\begin{figure*}
\centerline{\psfig{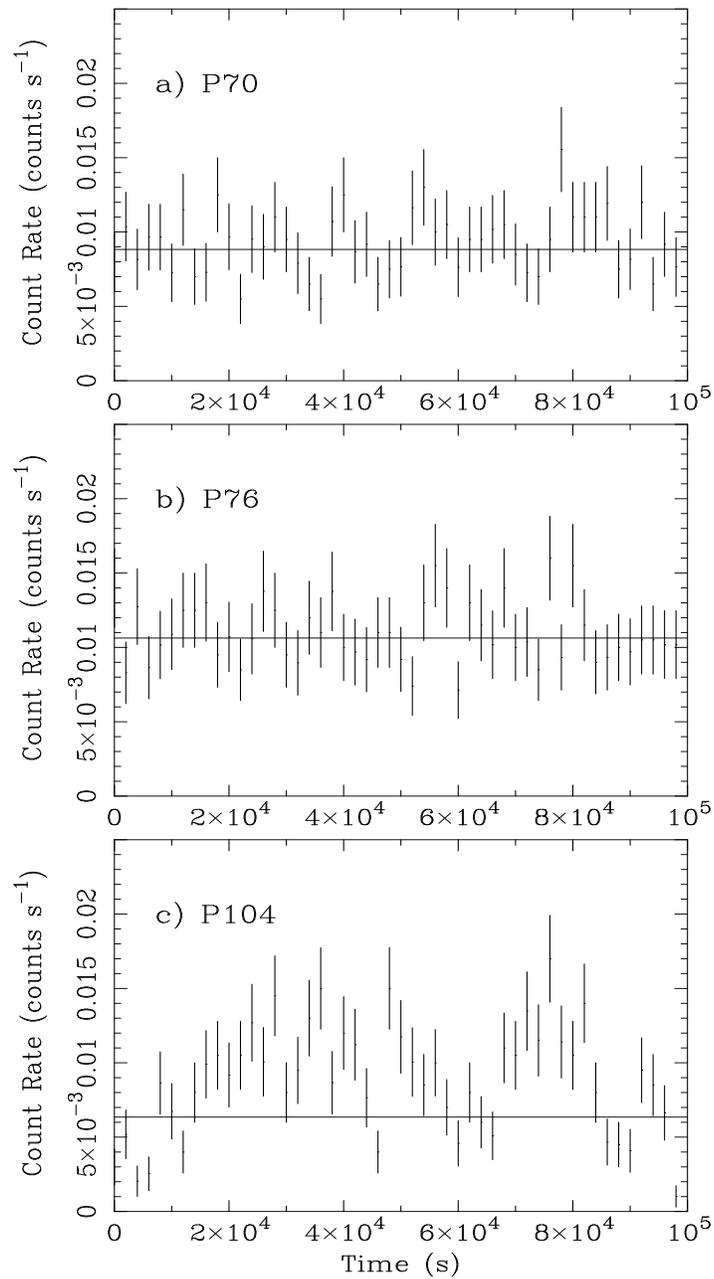}}
\caption{Light curves for the three hypernovae remnant
candidates, a) P70, b) P76, and c) P104.  The data were
binned into 2~ks intervals.}
\label{fig:lc}
\end{figure*}

P104 (MF83/H36), on the other hand, shows significant variation 
from a constant flux with a reduced $\chi^2$ value of 4.89, which 
clearly rules out a hypernova remnant origin.  The light curves are shown 
in Figure~\ref{fig:lc}.  We note that one of the two associations 
discussed by \citet{w99}, MF37/H19, which 
were outside of the region covered by the S3 chip in our observation was
also eliminated as a hypernova remnant candidate because of its temporal 
variability.

\begin{deluxetable}{cccccccccc}
\rotate
\tabletypesize{\scriptsize}
\tablecaption{Source Parameters - Power Law. 
\label{tbl:details-1}}
\tablewidth{0pt}
\tablehead{
\colhead{{\it Chandra}} & \colhead{Wang et al. (1999)}
& \multicolumn{8}{c}{This Work}  \\
\cline{3-10} \\
\colhead{Source}   & \colhead{S$_{14}$(0.5-2 keV)\tablenotemark{a,b}}
    & \colhead{Count Rate\tablenotemark{c}}
    & \colhead{S$_{14}$(0.5-2 keV)\tablenotemark{a}}
    & \colhead{S$_{14}$(0.5-10 keV)\tablenotemark{a}}
    & \colhead{L$_{38}$(0.5-10 keV)\tablenotemark{d,e}}
    & \colhead{$\Gamma$\tablenotemark{f}\ or {\it k}T\tablenotemark{g}}
    & \colhead{\NH\tablenotemark{h}}
    & \colhead{$\chi^2/\nu/P(\chi^2,\nu)\tablenotemark{i}$}
    & \colhead{Variable\tablenotemark{j}}}
\startdata
P67 (MF54)  & $-$   & 0.26  & 0.049 & $0.051\pm0.024$  & 0.043
  & $0.73\pm0.23$\tablenotemark{g} & 0.1\tablenotemark{k} & 1.82/2/0.48 & $-$ \\
P70         & 1.72  &  8.7  & 1.9   &  $7.5\pm1.2$  & 5.3
  & $1.65\pm0.13$\tablenotemark{f} & $1.7\pm0.4$ & 28.1/24/0.25 & No (0.92) \\
P76         & 1.61  & 10.1  & 2.0   & $11.0\pm1.9$  & 9.1
  & $1.75\pm0.12$\tablenotemark{f} & $5.4\pm0.7$ & 24.9/30/0.73 & No (0.76) \\
P104 (MF83) & 1.32  &  6.9  & 1.5   &  $2.3\pm0.3$  & 2.0
  & $2.88\pm0.24$\tablenotemark{f} & $1.4\pm0.4$ & 21.3/21/0.44 & Yes (4.89) \\
\enddata
\tablenotetext{a}{Units of $10^{-14}$ ergs cm$^{-2}$ s$^{-1}$.  The quoted
errors, when listed, are $1\sigma$.}
\tablenotetext{b}{Rederived from the \citet{wea99} {\it ROSAT} HRI count
rates using the {\it Chandra} fitted spectrum.}
\tablenotetext{c}{Units of $10^{-3}$ counts s$^{-1}$.}
\tablenotetext{d}{Units of $10^{38}$ ergs s$^{-1}$.}
\tablenotetext{e}{Deabsorbed luminosity.}
\tablenotetext{f}{Photon index from the power-law fit, the quoted errors
are $1\sigma$.}
\tablenotetext{g}{Temperature (keV) of fitted thermal equilibrium
(Raymond \& Smith 1977) model, the quoted errors are $1\sigma$.}
\tablenotetext{h}{Units of $10^{21}$ HI cm$^{-2}$, the quoted errors
are $1\sigma$.}
\tablenotetext{i}{$\chi^2$ value/degrees of freedom/$\chi^2_\nu$
probability.}
\tablenotetext{j}{Constant fit to data binned in 2.0~ks bins, $\chi^2_\nu$
value in parentheses.}
\tablenotetext{k}{Fixed value.}
\end{deluxetable}

The sources show only slight long-term variation in their fluxes when 
compared to the \citet{wea99} data.  We have recalculated the fluxes from 
the {\it ROSAT} HRI deep-exposure count rates using the 
{\it Chandra}-determined spectrum (see \S~\ref{sec:spec}) for the 
conversion, and the current values for P70, P76, and P104 are 10\%, 24\%,
and 14\% higher, respectively (Table~\ref{tbl:details-1}), where the 
uncertainties from counting statistics are $\sim10$\%.  The fact 
that the fluxes are all systematically higher for the three independent 
sources suggests that there may be an error in the interobservatory 
comparison.  With such a systematic error at the $\sim15$\% level, then
all sources would have varied at only the $\sim5$\% level.  We note that 
the flux of P104 in the {\it ROSAT} PSPC observation determined using the 
{\it Chandra} spectral fit is only $\sim4$\% higher than the current value.
 
\subsection{Spectral Properties}
\label{sec:spec}

\begin{figure*}
\centerline{\psfig{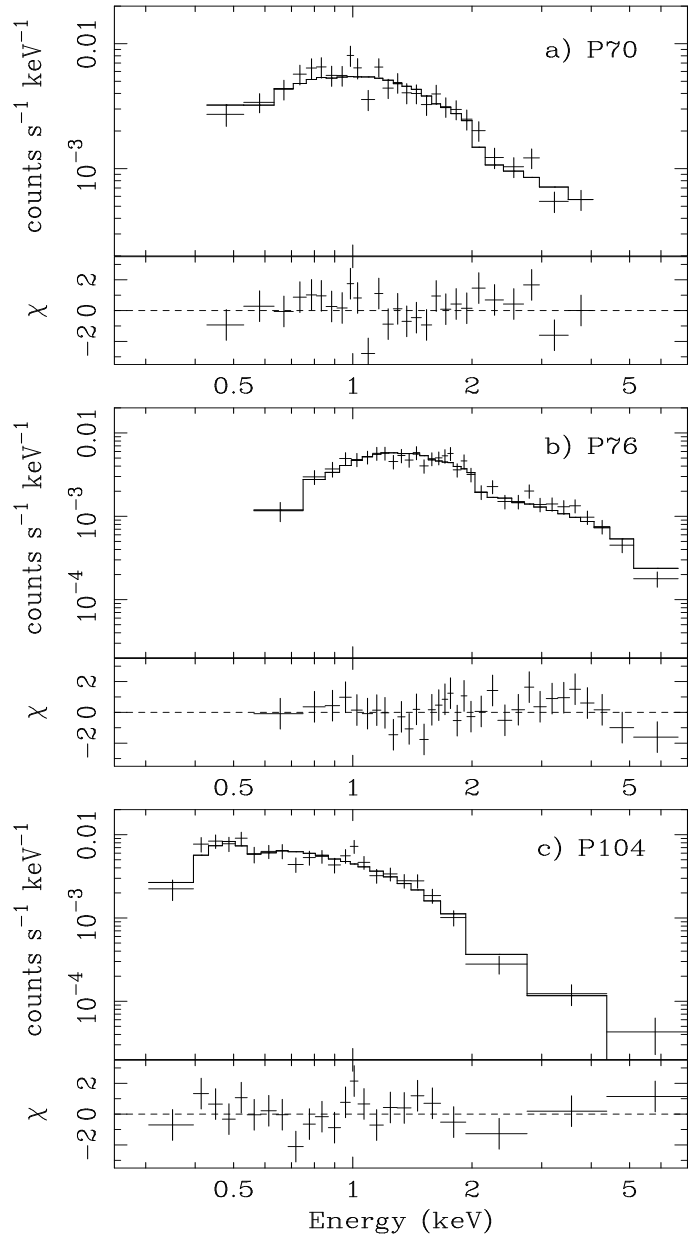}}
\caption{Fitted spectra of a) P70, b) P76,
and c) P104.  The upper panel for each individual plot shows
the data and the fitted power law model while the lower panel shows the
deviations from the model.}
\label{fig:spec}
\end{figure*}

\begin{deluxetable}{ccccccc}
\tabletypesize{\scriptsize}
\tablecaption{Source Parameters - Brem{\ss}trahlung. 
\label{tbl:details-2}}
\tablewidth{0pt}
\tablehead{
\colhead{Source}
    & \colhead{S$_{14}$(0.5-2 keV)\tablenotemark{a}}
    & \colhead{S$_{14}$(0.5-10 keV)\tablenotemark{a}}
    & \colhead{L$_{38}$(0.5-10 keV)\tablenotemark{b,c}}
    & \colhead{T (keV)\tablenotemark{d}}
    & \colhead{\NH\tablenotemark{e}}
    & \colhead{$\chi^2/\nu/P(\chi^2,\nu)$\tablenotemark{f}}}
\startdata
P70  & 1.9 &  $6.4\pm0.6$ & 4.5 & $7.7\pm3.2$   & $1.5\pm0.4$ & 27.6/24/0.28
\\
P76  & 2.0 & $10.0\pm1.0$ & 8.0 & $7.5\pm2.2$   & $4.6\pm0.6$ & 21.1/30/0.70
\\
P104 & 1.5 &  $1.9\pm0.5$ & 1.4 & $1.11\pm0.22$ & $0.7\pm0.2$ & 28.4/21/0.13
\\
\enddata
\tablenotetext{a}{Units of $10^{-14}$ ergs cm$^{-2}$ s$^{-1}$.  The quoted
errors, when listed, are $1\sigma$.}
\tablenotetext{b}{Units of $10^{38}$ ergs s$^{-1}$.}
\tablenotetext{c}{Deabsorbed luminosity.}
\tablenotetext{d}{Temperature of the brem{\ss}trahlung fit, the quoted
errors are $1\sigma$.}
\tablenotetext{e}{Units of $10^{21}$ HI cm$^{-2}$, the quoted errors
are $1\sigma$.}
\tablenotetext{f}{$\chi^2$ value/degrees of freedom/$\chi^2_\nu$
probability.}
\end{deluxetable}

The fitted spectra for P70, P76,  and P104 are 
shown in Figure~\ref{fig:spec}.  In all three cases a simple
absorbed power law spectrum adequately fits the data, as shown in 
Table~\ref{tbl:details-1}.  In all cases the power-law 
fits show evidence for additional absorption beyond that provided 
by the Milky Way, ranging from $\sim1-5\times10^{21}$ HI cm$^{-2}$.  The 
upper values are well above the typical thickness of the M101 disk implying 
that the sources are embedded in significant enhancements in the density of 
the interstellar medium, or are sources with intrinsic absorption.  A 
brem{\ss}trahlung model (Table~\ref{tbl:details-2}) fit the harder spectra 
of P70 and P76 well but was less successful for the softer spectrum of P104.  
The brem{\ss}trahlung fits require lower values for \NH, and except for 
P76 are consistent with the Galactic column plus a reasonable value for 
the absorption by the interstellar medium in the disk of M101. 

The spectra of P70 and P76 are reasonably well fit by thermal spectra 
with $kT\sim8-10$~keV, which are essentially power-law in shape.  The
spectrum of P104 is poorly fit by a thermal emission model, although 
the best-fit temperature is significantly lower at $\sim3$~keV.  Thus,
in all three cases, the source spectra are best fit by continuum emission 
models.

\subsection{Positions}

\begin{figure*}[t!]
\centerline{\psfig{figure=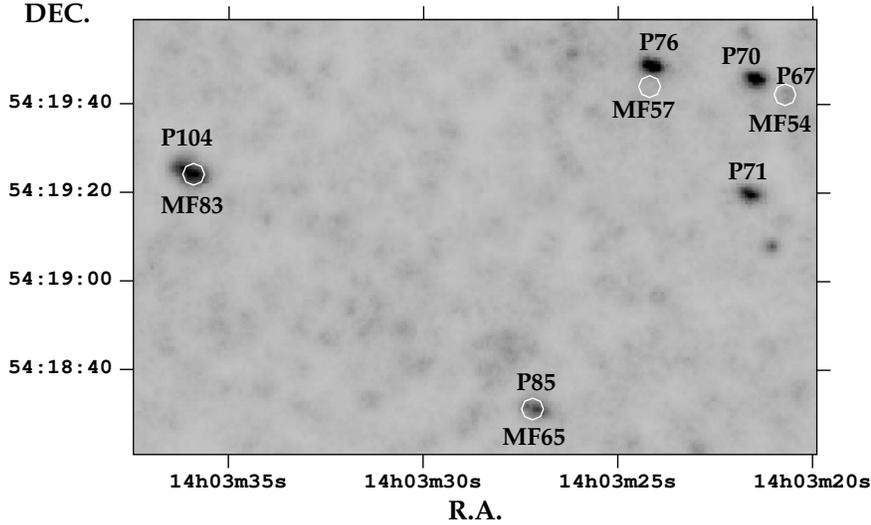,width=12.0cm}}
\caption{Image of the {\it Chandra} data in the
neighborhood of the three sources, P70, P76, and P104. The \citet{mf97}
locations of four supernova remnants are marked with $5''$ circles.
For clarity, the data have been smoothed using an adaptive filter with
a kernel of 9 counts.}
\label{fig:image}
\end{figure*}

{\it Chandra} source coordinates \citep{pea01} were determined using 
reprocessed data providing positions accurate to $\sim1''$.  
The \citet{pea01} and \citet{w99} positions of the three 
sources considered here agree to $\lsim3''$.
Figure~\ref{fig:image} shows the {\it Chandra} data for a limited region of 
M101 surrounding the three hypernova remnant candidates.  The positions of 
the \citet{mf97} 
supernova remnants are indicated by $5''$ diameter circles.  
The positions of MF54 and MF57 in the upper right of the image are clearly 
inconsistent with the bright X-ray sources in the vicinity with offsets of 
$8\farcs3$ and $4\farcs8$, respectively.  (There {\it is} an apparent
faint enhancement at the source position of MF54, the source at 
the far right, which is discussed below.)  However, the positions are close 
enough to have 
justified the conclusion, drawn by \citet{w99} from the lower angular 
resolution {\it ROSAT} HRI data, that they were possibly associated, 
and are actually in relatively good agreement with the offsets of $6\farcs2$ 
and $4\farcs0$ quoted in that paper.  The projected separations between the 
X-ray positions of P70 and P76 and the optical positions of MF54 and 
MF57 are 290 pc and 170 pc, respectively.

\subsection{MF54, Will the Real SNR Please Stand Up}

We have detected a faint source (P67) at the optically-determined 
position of MF54 at the $\sim4\sigma$ level.  Accepting the lower 
reliability of spectral fits to data with poor statistics (note 
that there are $\sim25$ source counts and $\sim40$ background 
counts in the spectrum from the source region), 
we have extracted and fit the data to test for consistency with the 
attribution of the observed emission to an SNR source.  Using an absorbed 
thermal model, the best fit yields ${\rm kT}=0.54$~keV and 
$\NH=6.3\times10^{21}$~\hi~cm$^{-2}$ with $\chi^2_\nu=0.19$ (with one 
degree of freedom).  The fitted values for $kT$ and $\NH$  are essentially 
unconstrained, however.  The implied flux  ($0.5-10.0$~keV) is 
$6.0\times10^{-16}$~ergs~cm$^{-2}$~s$^{-1}$ with an unabsorbed
luminosity of $2.4\times10^{37}$~ergs~s$^{-1}$.  Fixing the absorption 
at $10^{21}$~\hi~cm$^{-2}$, a level closer to the nominal disk 
absorption of M101 plus that from the Milky Way, and refitting the 
data yields ${\rm kT}=0.73\pm0.23$~keV with $\chi^2_\nu=0.62$ (with 
two degrees of freedom).  The flux and unabsorbed luminosity in this 
case are $5.1\times10^{-16}$~ergs~cm$^{-2}$~s$^{-1}$ and 
$4.3\times10^{36}$~ergs~s$^{-1}$.  The latter values 
are completely reasonable for a supernova remnant with a diameter 
of $\sim20$~pc (determined by \citet{mf97}), and the quality of the 
fit is still reasonable.

Alternatively we fit continuum power law and blackbody spectra to the 
data.  The best fit for an absorbed power law, with $\chi^2_\nu=2.96$, 
required an exceptionally soft photon index of 8.0 and absorption of 
$\sim1.2\times10^{22}$~\hi~cm$^{-2}$.  The blackbody spectrum fared
slightly worse with ${\rm kT}=0.20$~keV and 
$\NH=3.4\times10^{21}$~\hi~cm$^{-2}$ with $\chi^2_\nu=3.29$ (as with
the thermal model, the values for the fitted parameters were essentially 
unconstrained).  While not 
exceptionally significant (even the black body fit was still acceptable 
at the $\sim7$\% level), these poorer quality fits support the 
attribution of the detected flux to thermal emission from a 
supernova remnant.  In both cases the fitted soft spectra coupled 
with fitted high absorption act to imitate the intrinsically more narrowly 
peaked spectrum of the probable thermal emission.

\section{Discussion}
\label{sec:discussion}

If the bright sources are not hypernovae remnants, 
what are they?  In the case of
MF83, where the X-ray emission is coincident with the optical position
of the supernova remnant, a recent paper by \citet{lea01} provides some
information.  Based on deep ground-based and HST optical images and
echelle spectra, they provided a detailed look at MF83, and found 
that it has a diameter of $\sim270$~pc (rather large for a 
supernova remnant), contains several OB associations, and has several 
\hii\ regions associated with its rim.  Because of these aspects as well 
as arguments based on the observed expansion velocities and optical 
line ratios, \citet{lea01} suggest that MF83 should be more accurately
characterized as a superbubble and star-formation region rather than as a 
supernova remnant.  With the temporal variability, power-law spectrum,
and unresolved nature of the {\it Chandra} source, we suggest that the
bulk of the observed flux is from an X-ray binary associated with the 
superbubble.

The bright sources associated by \citet{w99} with MF54 and MF57
provide a different challenge.  By their positions they are clearly not
associated with the supernova remnants nor with any objects in the 
Simbad catalog.  Given their power law spectra and intrinsic absorption,
either an AGN or X-ray binary origin are possible.  Using the 
$N(>S)$ curve published in \citet{mea00} (relying in the appropriate 
flux range on {\it ROSAT} data from \citet{hea98}), only $\sim0.7$ such 
sources should be found on an S3 chip observation at the flux level of 
$2\times10^{-14}$~ergs~cm$^{-2}$~s$^{-1}$ and above.  Finding two such
sources within $30''$ of each other is unlikely.  On the other hand, the
sources are $\sim4$~kpc from the galactic nucleus, about $\sim25$\% 
of the equivalent solar circle in M101.  Thus the identification of the
sources as X-ray binaries associated with M101 is not unreasonable.

\begin{figure}[!t]
\centerline{\psfig{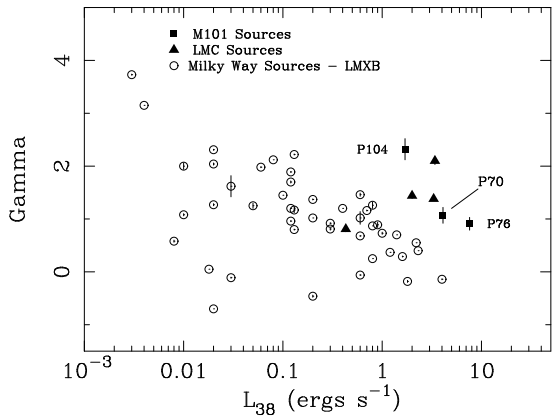}}
\caption{Plot of \citet{cs97} LMXB and current data.  The
X axis is the luminosity in units of $10^{38}$ ergs s$^{-1}$.  The Y axis
is the fitted value for the power law index $\Gamma$ in the USC model
(an absorbed power law with an exponential).}
\label{fig:lmxb}
\end{figure}

To further investigate the source identifications we compare the spectral
properties derived here to the sources in \citet{cs97} (their Figure~25),
where the value for $\Gamma$ from the fits to the unsaturated Compton
(USC) emission model are plotted as a function of luminosity.  Due to the
limited statistics at high energies, we were unable to constrain the scaling
energy of the USC model; we therefore adopted a typical Milky Way value of
4 keV.  Figure~\ref{fig:lmxb} displays the results, where we have also added
points for LMC X-1, X-3, and X-4.  The M101 sources occupy a region of this
diagram distinct from the Milky Way sources, but compare well with the LMC
sources.  One possible cause of the difference is the nature of the compact
object: while most of the Milky Way sources discussed by \citet{cs97} are
neutron star binaries, LMC X-1 and X-3 are black hole candidates.  Since
the M101 sources are all near or above the Eddington limit for a neutron
star binary, they are black hole candidates as well.  The separation seen in
Figure~\ref{fig:lmxb} can be taken as an additional support for the view that
the M101 sources are black hole binaries in a soft, high state.  The spectral
properties of these and other bright sources of M101 will be discussed in
\citet{mea01}.

\section{Conclusions}
\label{sec:results}

We have considered the three X-ray source/optical supernova remnant 
associations that were suggested as possible hypernova remnant candidates 
by \citet{w99}, which were covered by our {\it Chandra} deep exposure of 
M101.  The most plausible case, that of MF83, is clearly ruled out because
of its temporal variability.  The two less plausible candidates, 
MF54 and MF57, are ruled out by their positions, which are inconsistent with 
the optically identified supernova remnants.  However, we have detected an 
X-ray source consistent with the optical position of MF54 which has a 
spectrum and luminosity consistent with those of a normal supernova remnant.
The bright sources, P70, P76, and P104, have luminosities on the high 
end of the Milky Way LMXB distribution but have spectra which are 
significantly softer.  However, the attribution of the observed emission 
to LMXBs is not unreasonable.

The positional correlation, at least in the case of P104/MF83, of the X-ray 
sources with the optically-identified supernova remnants is suggestive of 
an association between the two, even if the emission is not extended and 
from the remnant itself.  Clearly, supernovae provide a simple mechanism 
for producing a black hole as well as the remnant.  However, for the P70/MF54 
and P76/MF57 pairs the situation is not so simple.  The projected separations  
between the binary systems and the centers of the supernova remnants are on the 
order of 200 pc, which is ten times the optical diameters of the remnants.
Thus while the progenitor(s) of the P104 binary system could easily have 
contributed to the energy of the MF83 superbubble, it is much less likely 
that progenitors of P70 and P76 contributed similarly to MF54 and MF57.

\acknowledgements

We would like to thank the referees for pointing out several significant 
shortcomings in the initial draft of this paper, and for several additional 
comments which led to its improvement.  We also acknowledge the support of 
{\it Chandra} grant 01900441.



\begin{thebibliography}{}
\bibitem[Bautz et al.(1998)]{bea98} Bautz, M. W. et al. 1998, 
    Proc. SPIE, 3444, 210
\bibitem[Christian \& Swank(1997)]{cs97} Christian, D. J., \& Swank, J. H. 
    1997, \apjs, 109, 177
\bibitem[Fryer \& Woosley(1998)]{fw98} Fryer, C. L., \& Woosley, S. E. 1998
    ApJL, 502, L9
\bibitem[Garmire et al.(1992)]{gea92} Garmire, G. P., Ricker, G. R., 
    Bautz, M. W., Burke, B., Burrows, D. N., Collins, S. A., Doty, J. P., 
    Gendreau, K., \& Lumb, D. H. 1992, in AIAA, Space Programs and 
    Technologies Conf. (New York: AIAA), 8
 Nousek, J. A.
\bibitem[Hasinger et al.(1998)]{hea98} Hasinger, G., Burg, R., 
    Giacconi, R., Schmidt, M., Tr\"umper, J., \& Zamorani, G. 1998, A\&A,
    329, 482
\bibitem[Lai et al.(2001)]{lea01} Lai, S.-P., Chu, Y.-H., Chen, C.-H. R.,
    Ciardullo, R., \& Grebel, E. K., submitted
\bibitem[Matonick \& Fesen(1997)]{mf97} Matonick, D. M., \& Fesen, R. A. 
    1997, ApJS, 112, 49
\bibitem[Mukai et al.(2001)]{mea01} Mukai, K., et al. 2001, \apj, in
    preparation
\bibitem[Mushotzky et al.(2000)]{mea00} Mushotzky, R. F., Cowie, L. L.,
    Barger, A. J., \& Arnaud, K. A. 2000, Nat, 404, 459
\bibitem[Paczy\'nski(1998)]{p98} Paczy\'nski, B. 1998, ApJ, 494, L45
\bibitem[Pence et al.(2001)]{pea01} Pence, W., et al., 2001, \apj, in
    preparation
\bibitem[Snowden \& Pietsch(1995)]{sp95} Snowden, S. L., \& Pietsch, W. 
    1955, \apj, 452, 627
\bibitem[Stetson et al.(1998)]{sea98} Stetson, P. B., et al. 1998, ApJ,
    290, 449
\bibitem[Wang(1999)]{w99} Wang, Q. D. 1999, ApJL, 517, L27
\bibitem[Wang, Immler, \& Pietsch(1999)]{wea99} Wang, Q. D., Immler, S., 
    \& Pietsch, W. 1999, ApJ, 523, 121
\bibitem[Weisskopf, O'Dell, \& van Speybroeck(1996)]{wos96} Weisskopf, M. 
    C., O'dell, S. L., \& van Speybroeck, L. P. 1996, Proc. SPIE, 2805, 2
\end{thebibliography}
\end{document}